# Derivation of Elastic Wave Equation from New Motion Description

Peng SHI（时鹏）[1, 2, *]

[1] Department of Astronautics and Mechanics, Harbin Institute of Technology, Postbox 344, 92 West Dazhi Street, Harbin, 150001, P. R. China

[2] Technology Centre of China Petroleum Logging Co., LTD., CNPC Well Logging Laboratory, 50 Zhangba five road, Xi'an, 710077, P. R. China

[*] Corresponding author: sp198911@outlook.com;

*Abstract.* In classical mechanics, the motion of an object is described with Newton's three laws of motion, which means that the motion of the material elements composing a continuum can be described with the particle model. However, this viewpoint is not objective, since the existence of transverse wave cannot be predicted by the theory of elasticity based on the particle model. In this paper, the material element of an elastomer is regarded as a rigid body, and the traditional elastic wave equation is derived based on it. In the derivation, the constitutive relations and strain-displacement relations are correspondingly modified. The study reveals that the longitudinal and transverse waves in elastomer correspond to the translational and rotational motion of the material element, respectively. Besides, the reciprocity of shear stress and shear strain is no longer requisite in continuum mechanics, and the local rigid body rotation contributes stress.

## 1. Introduction

Classical mechanics deals with the motion of an object under various forces in the absolute concept of time-space [1-3]. To describe the mechanical properties of different kind of objects under the action of external forces, various branches of classical mechanics have been proposed and developed, such as particle mechanics, rigid body mechanics and continuum mechanics [4-8]. With the efforts of mechanics masters, classical mechanics has been widely used in natural science and engineering technology [9, 10]. Due to the success in describing the motion of an object with particle model, the community of mechanics generally believes that the basic laws of classical mechanics are Newton's three laws of motion or other mechanical principles related to and equivalent to them [1-11]. However, this is far from objective. This study shows that the motion of a continuum in the domain of continuum mechanics does not satisfy Newton's three laws of motion.

Continuum mechanics, as an important branch of classical mechanics, concerns with the stresses in solids and fluids and the deformation or flow of these materials [7, 8]. In continuum mechanics, the macroscopic medium is considered to satisfy the continuum hypothesis, where the structures of real fluid and solid are considered to be perfectly continuous and are paid no attention to their molecular structure. The nonzero-volume elements constituting a continuous media are treated as a particle and their dynamics is described with Newton's second law under the initial configuration [7, 8, 10]. On this basis, the classical theory of continuum mechanics is established. To make the motion of material element satisfy the particle model, the Cauchy stress tensor showing the stress state at a point under the current configuration is a second-order symmetric stress tensor [10, 12]. For the sake of ensuring the symmetry of stress tensor the strain and strain rate tensors, considered to contribute the stress and described with the gradient of displacement and velocity respectively, are also second-order symmetric tensors [12, 13]. In the continuum mechanics theory, there exists a fatal conflict that the rotation of the material element is admitted when we describe the deformation of a continuous material but its rotation is ignored when we describe its motion [7, 8, 12, 13]. This conflict makes it difficult to explain why the displacement in the displacement equations of motion and the velocity in Navier-Stokes equations are curl, since the displacement and velocity caused by strain and strain rate respectively should be irrotational. In my opinion, the problems in classical theory of continuum mechanics are caused by the non-universality of Newton's laws of motion. In recent decades, some new continuum mechanics models have been proposed and the rotation of material elements has been declared [14, 15]. However, no researcher declares that these models violate Newton's three laws of motion.

In this study, I show that Newton's three laws of motion are not the basic laws of the object motion via taking the derivation of elastic wave equation as an example. The remainder of the paper is organized as follows. First, whether the motion of rigid body can be regarded as the motion of particle system is discussed, and an error is pointed out in the derivation of elastic wave equation, in which the transverse wave (or shear wave) equation should not have been obtained. Then, the elastic wave equations are derived by treating the material element as a rigid body whose translation and rotation should be considered. Finally, a new differential motion equation of the continuum is obtained by assuming that the rigid body model can describe the motion of its constituent material element.

2. **Problem statement**

Particle and rigid body are two basic ideal model in classical mechanics [1-3]. A particle is defined as a point with mass but no volume or shape to replace an object without considering its volume and shape. A rigid body is defined as an object for whom the change of shape and size can be ignored under the action of forces. At present, researchers generally believe that the rotation of a rigid body is the set of motion of a particle system, since the moment of momentum of a rigid body rotating around an axis can be described by the motion of particles around the axis [1-6]. In fact, this view is wrong on two levels. Level one, a particle system can't form a rigid body owing to that the particle has no volume. That is, the particle mention above is a rigid body higher order infinitesimal in size, whose rotation can be ignored. Level two, the rotation of the rigid bodies, which are called particles above, can't be ignored, at least not all or all the time. For example, when a rigid body rotates around its center of mass, the rotation of the constituent rigid bodies whose centroid are on and near the axis of rotation of the macroscopic rigid body can't be ignored. Therefore, the particle model is not needed in classical mechanics and can be replaced by the rigid body model due to the volume of any object with finite density is nonzero. The above analysis shows whether the rotation of an object can be ignored is dependent on the problem to solve, not its size.

In classical continuum mechanics, a continuum is regarded as a set of particles. The fundamental law of the dynamics applied to a part of the continuous medium is written in the following form [7, 8]:

$$\frac{\mathrm{D}}{\mathrm{D}t}\oint \rho v \mathrm{d}V - \oint \boldsymbol{T} \cdot \mathrm{d}\boldsymbol{S} - \oint \boldsymbol{f} \mathrm{d}V = 0, \tag{1}$$

where D/Dt is the material derivative, $v$ is the speed of the material element, $\rho$ is the mass density, $V$ is the volume of the continuous medium, $T$ is the surface force, $S$ is the surface element vector, $f$ is the body force which is an irrotational field. Treating a material element as particle and only considering its translation, the equation of motion in differential form for the continuous medium with small deformation is expressed in the following form [12]:

$$\rho \frac{\partial^2 \boldsymbol{u}}{\partial t^2} - \nabla \cdot \boldsymbol{\sigma}^S - \boldsymbol{f} = 0, \tag{2}$$

$$\boldsymbol{T} = \boldsymbol{\sigma}^S \cdot \boldsymbol{n}, \tag{3}$$

here, $\partial/\partial t$ is the time derivative, $\boldsymbol{u}$ is the displacement, $\nabla$ is the vector operator del, $\boldsymbol{\sigma}^S$ is the symmetric second-order stress tensor, $\boldsymbol{n}$ is the unit vector of outer normal of surface element. Equation (2) is derived with the Gauss flux theorem, which describes the motion of the material element under the action of force at a certain position.

When the continuous medium is an isotropic elastomer, its constitutive relations and strain-displacement relations in component form are expressed as [12, 16]

$$\sigma_{ij}^S = C_{ijkl} e_{kl}^S, \tag{4}$$

$$e_{kl}^S = \tfrac{1}{2}\left(u_{k,l} + u_{l,k}\right), \tag{5}$$

$$C_{ijkl} = \lambda \delta_{ij} \delta_{kl} + \mu \left(\delta_{ik}\delta_{jl} + \delta_{il}\delta_{jk}\right), \tag{6}$$

where, $e_{kl}^S$ is strain tensor, $C_{ijkl}$ is elastic tensor, $\delta_{ij}$ is the Kronecker delta. $\lambda$ and $\mu$ are the Lamé constants. If the strain-displacement relations are substituted into the constitutive relations and the expressions for the stresses are subsequently substituted in the equation of motion, the displacement equations of motion can be obtained in vector notation [12, 16]:

$$\mu \nabla^2 \boldsymbol{u} + (\lambda + \mu)\nabla \nabla \cdot \boldsymbol{u} + \boldsymbol{f} = \rho \frac{\partial^2 \boldsymbol{u}}{\partial t^2}. \tag{7}$$

It should be pointed out that the displacement in Equation (7) should only include the displacement caused by deformation, because the rotation of local rigid body does not contribute to the stress. Therefore, the displacement in Equation (7) should be irrotational, and the following formula is true:

$$\nabla^2 \boldsymbol{u} = \nabla \nabla \cdot \boldsymbol{u}. \tag{8}$$

Substituting Equation (8) in to Equation (7), the following formula is obtained:

$$(\lambda + 2\mu)\nabla \nabla \cdot \boldsymbol{u} = \rho \frac{\partial^2 \boldsymbol{u}}{\partial t^2}. \tag{9}$$

It is seen from Equation (9) that the longitudinal wave equation is only derived. The existence of transverse waves can't be predicted by the elastic theory which uses the particle model to describe the motion of the material element. The conclusion can be also obtained by describing the motion equation in the orthogonal coordinate system whose directions of coordinate axes are parallel to the directions of three principal stresses. In the orthogonal coordinate system, only normal stresses are nonzero in the equation of motion. Correspondingly, only the normal strains are nonzero. Hence, the transverse wave equation can't be obtain.

The difference of the derivation of elastic wave equation between the study and former ones is from Equation (8). In the traditional derivation process of elastic wave equation, the hypothesis that the rotation of local rigid body does not contribute to the stress is broken, and the stresses contributed by local rigid body rotation is introduced with the formula [16]:

$$\nabla^2 \boldsymbol{u} = \nabla(\nabla \cdot \boldsymbol{u}) - \nabla \times \nabla \times \boldsymbol{u}. \tag{10}$$

Then, the elastic wave equation including the transverse wave is obtained by submitting Equation (10) in to Equation (7). This indicates that the motion of the material elements composing an elastomer may satisfy the rigid body model, but the particle model.

## 3  Wave equation derivation from rigid body model

From the derivation of elastic wave equation by considering the material element as a particle, it is known that the longitudinal waves correspond to the translation of material elements. Under the linear elastic hypothesis, the deformation and motion for the continuous material satisfy the superposition principle. Below, I only need to derive the displacement equations of motion from the angular momentum conservation of rigid body rotation.

The angular momentum conservation of a part of the continuous medium should be expressed as:

$$\int -\boldsymbol{R} \times \frac{1}{2}(\boldsymbol{\varepsilon}:\boldsymbol{\sigma}) \mathrm{d}l = \int \boldsymbol{R} \times \rho \frac{\partial^2 \boldsymbol{u}}{\partial t^2} \mathrm{d}S , \qquad (11)$$

$$\boldsymbol{\varepsilon}:\boldsymbol{\sigma} = \boldsymbol{\varepsilon}:\boldsymbol{\sigma}^A = \frac{1}{2}\boldsymbol{\varepsilon}:\left(\boldsymbol{\sigma} - \boldsymbol{\sigma}^T\right), \qquad (12)$$

where, $\boldsymbol{R}$ is the radius vector of the origin at the centroid, $\boldsymbol{\sigma}$ is the stress tensor which is asymmetrical, $\boldsymbol{\sigma}^T$ and $\boldsymbol{\sigma}^A$ are the transposition and antisymmetric part of $\boldsymbol{\sigma}$, respectively, $\boldsymbol{\varepsilon}$ is the permutation symbol, $\boldsymbol{l}$ is the boundary line vector, $\boldsymbol{S}$ is the surface element vector. With the Stokes' theorem, the conservation of angular momentum of material element can be expressed in differential form as:

$$-\frac{1}{2}\nabla \times (\boldsymbol{\varepsilon}:\boldsymbol{\sigma}) = \rho \frac{\partial^2 \boldsymbol{u}}{\partial t^2} . \qquad (13)$$

The classical continuum mechanics doesn't consider the material element rotation. Therefore, the shear stresses are reciprocal. The strain tensor in the classical theory of elasticity, which describes the deformation of elastomer under the action of force, must also satisfy the shear strain reciprocity. The symmetric part of the displacement gradient is used to express the strain, and the anti-symmetric part of the displacement gradient is treated as the local rigid body rotation without contributing stress [12]. When the motion of material element is treated as a rigid body, the shear stress reciprocity need to be broken, in order to balance the moment of momentum related to the local rigid body rotation. Correspondingly, the local rigid body rotation need to contribute to stress, and the stress field produced by local rigid body rotation are pure curl field. It needs to be pointed out that the local rigid body rotation does not include the rigid-body rotation of the continuum. In the following statement, the continuum is assumed not to rotate as a whole.

Under the new motion description, the constitutive relations need to be modified appropriately to have the deformation of an elastomer produces an asymmetrical strain tensor. The study believes that the stress-strain relationships and strain-displacement relationships of elastomer should be expressed as follows:

$$\boldsymbol{\sigma} = \boldsymbol{C} : \boldsymbol{\xi}, \tag{14}$$

$$\boldsymbol{\xi} = \nabla \boldsymbol{u}, \tag{15}$$

here $\boldsymbol{\xi}$ is the strain including the traditional defined strain and local rigid body rotation. For isotropic elastomers, the elastic tensor should be expressed as

$$C_{ijkl} = \lambda \delta_{ij} \delta_{kl} + 2\mu \delta_{ik} \delta_{jl}. \tag{16}$$

It is seen from Equations (14) to (16) that the stress tensor is asymmetric when the displacement field is rotational, and symmetric when the displacement field is irrotational. This means that Equations (14) to (16) contain the case in the classical theory of elasticity. In fact, the traditional representation of the elastic tensor is redundant since that the local rigid body rotation does not contribute to the stress has been declared in the theory of elasticity. Substituting Equations (14) and (15) into Eq. (13), the following equation can be

derived:

$$-\nabla \times \left(\frac{1}{2}\boldsymbol{\varepsilon} : 2\mu\boldsymbol{\Omega}\right) = \rho \frac{\partial^2 \boldsymbol{u}}{\partial t^2}, \qquad (17)$$

where

$$\boldsymbol{\Omega} = \frac{1}{2}\left(\nabla \boldsymbol{u} - \nabla \boldsymbol{u}^T\right), \qquad (18)$$

$\boldsymbol{\Omega}$ is an anti-symmetric tensor representing the strength of local rigid body rotation. Due to the following Equation exists [17]

$$\nabla \times \boldsymbol{u} = \boldsymbol{\varepsilon} : \boldsymbol{\Omega}, \qquad (19)$$

Equation (17) can be rewritten as:

$$-\mu \nabla \times \nabla \times \boldsymbol{u} = \rho \frac{\partial^2 \boldsymbol{u}}{\partial t^2}. \qquad (20)$$

Equation (20) is the transverse wave equation. By adding Equation (9) and Equation (20), the elastic wave equation can be obtained:

$$(\lambda + \mu)\nabla\nabla \cdot \boldsymbol{u} - \mu \nabla \times \nabla \times \boldsymbol{u} + \boldsymbol{f} = \rho \frac{\partial^2 \boldsymbol{u}}{\partial t^2}. \qquad (21)$$

The wave equation is identical to the traditional one [12, 16]. This means that the motion of elastomer can be described with the rigid body model. In the classical theory of elasticity, though the local rigid body rotation is admitted, the traditionally defined deformation is only considered for the deformation coordination, which means that the material elements composing an elastomer can rotate freely. This is inconsistent with the facts. In admitting the rotation of material element, the deformation coordination should be described as follows [18]:

$$\xi_{ij,k} = \xi_{kj,i} = u_{j,ik}. \qquad (22)$$

Based on the above analysis, it is obtained that the motion of an elastomer should be governed with Equations (2) and (13), its stress-strain relationships and strain-displacement relationships are governed with Equations (14) and (15), respectively, and its deformation coordination is governed with Equations (22).

From the above analysis it has been concluded that the particle model cannot fully describe the motion of the material elements composing a continuum. Since the material elements of an elastomer can be regarded as the rigid body in the description of motion, it is reasonable to assume that the motion of the material elements composing a continuum can be described with the rigid body model. Under the assumption, the differential equation of motion for a continuum should be expressed as

$$\frac{\mathrm{D}(\rho \boldsymbol{v}^S)}{\mathrm{D}t} + \rho \frac{\partial \boldsymbol{v}^A}{\partial t} = \nabla \cdot \boldsymbol{\sigma} + \boldsymbol{f} - \frac{1}{2}\nabla \times (\boldsymbol{\varepsilon} : \boldsymbol{\sigma}) \ , \tag{23}$$

here, $v^S$ and $v^A$ are the velocity fields in continuum corresponding to the translation and rigid body rotation of material element, respectively. It is easily obtained that for the one-dimensional flow the velocity equation of motion derived from Equation (23) is the same as the Navier-Stokes equation. In this perspective, Equation (23) is successful in describing the motion of continuous medium.

## 4  Conclusion

The study has demonstrated the non-universality of Newton's three laws of motion via the derivation of elastic wave equation. The derivation reveals that only the longitudinal wave equation can be derived by treating the material elements composing elastomer as particles. To obtain the traditional elastic wave equation, the material elements need to be treated as rigid bodies when their motion is described. The derivation further reveals that the longitudinal and transverse waves in elastomer correspond to the translation and rigid body rotation of material element, respectively. Under the new motion description, a new theory of elasticity is established. The new theory shows that the reciprocity of shear stress and shear strain is not requisite in continuum mechanics, and the local rigid body rotation produces stress.

**Acknowledgments**

The author would like to thank WN Zou from Nanchang University, and HS Hu and W Guan from Harbin institute of technology for sharing their ideas. This research did not receive any specific grant from funding agencies in the public, commercial, or not-for profit sectors.